# Enhancing Recommendation Systems with GNNs and Addressing Over-Smoothing

Wenyi Liu
Independent Researcher
Nanjing, China

Ziqi Zhang
Independent Researcher
Ann Arbor, USA

Xinshi Li
Montclair State University
New Jersey, USA

Jiacheng Hu
Tulane University
New Orleans, USA

Yuanshuai Luo
Southwest Jiaotong University
Chengdu, China

Junliang Du *
Shanghai Jiao Tong University
Shanghai, China

*Abstract*—**This paper addresses key challenges in enhancing recommendation systems by leveraging Graph Neural Networks (GNNs) and addressing inherent limitations such as over-smoothing, which reduces model effectiveness as network hierarchy deepens. The proposed approach introduces three GNN-based recommendation models, specifically designed to mitigate over-smoothing through innovative mechanisms like residual connections and identity mapping within the aggregation propagation process. These modifications enable more effective information flow across layers, preserving essential user-item interaction details to improve recommendation accuracy. Additionally, the study emphasizes the critical need for interpretability in recommendation systems, aiming to provide transparent and justifiable suggestions tailored to dynamic user preferences. By integrating collaborative filtering with GNN architectures, the proposed models not only enhance predictive accuracy but also align recommendations more closely with individual behaviors, adapting to nuanced shifts in user interests. This work advances the field by tackling both technical and user-centric challenges, contributing to the development of robust and explainable recommendation systems capable of managing the complexity and scale of modern online environments.**

## I. INTRODUCTION

The rapid development of the online shopping and social media industries has made recommendation systems one of the essential tools on the internet. Recommendations for purchases on shopping platforms, songs on music platforms, and friends on social platforms are all applications of recommendation systems. Accurately modeling user preferences based on behaviors such as clicks, views, reads, and purchases is the core of an effective recommendation system. In the current era of information overload, recommendation models aim to recommend items of interest to users, addressing the issue of information overload. Graph Neural Networks (GNNs) [1] are widely applied in many predictive and recommendation tasks due to their ability to feature structured and collaborative recommendations of items and features.

However, as the network hierarchy increases, Graph Neural Networks suffer from severe over-smoothing issues that affect recommendation performance. At the same time, most current recommendation models adopt deep learning paradigms, which raise an urgent demand for the interpretability of recommendations. The focus of interpretable recommendations is not only to produce accurate recommendation results but also to provide convincing explanations of how and why a particular item is recommended to a specific user. Existing review-based interpretable recommendation methods mostly use static independent methods to extract potential feature representations from user and item reviews, representing user preferences as static feature vectors. However, users typically exhibit different preferences when interacting with different items. Therefore, around the above key issues and in light of the current models' shortcomings, three more efficient recommendation models are proposed, as follows:

In response to the severe over-smoothing issue in Graph Neural Networks, a collaborative filtering recommendation algorithm based on GNNs is proposed to improve recommendation performance. Building upon the use of GNNs, the mechanism of collaborative filtering (CF) [2] is added to address the smoothing problem faced by existing recommendation algorithms using GNNs. The algorithm introduces initial residual connections and identity mapping in the aggregation propagation process of building the network, ensuring that the network can perform deep learning and avoid the over-smoothing phenomenon that occurs after multiple convolutions in GNNs, thereby improving recommendation accuracy.

## II. RELATED WORK

The Graph Neural Networks (GNNs) are essential for modeling complex user-item interactions in recommendation systems, but over-smoothing remains a challenge. Techniques such as residual connections and improved aggregation strategies have been proposed to maintain feature distinction in deeper layers[3], supporting this work's focus on addressing this issue [4].

Collaborative filtering (CF), a cornerstone in recommendation systems, has been augmented by multimodal approaches. A. Liang [5] introduces frameworks leveraging multimodal transformers to enhance recommendations in cross-domain and personalized scenarios. These works highlight the importance of fusing diverse data modalities to adapt to dynamic user preferences.

Interpretability is vital for recommendation systems, and methods using transformer models [6]and self-supervised learning [7] have been developed to create transparent, explainable predictions [8]. These approaches align closely with the need for explainable GNN-based models.

Temporal and sequential data modeling methods, such as transforming time-series into interpretable sequences, inform this work's strategy for adaptive recommendations [9]. Techniques for capturing temporal dependencies ensure systems are responsive to user preference shifts [10].

Broader advancements in deep learning, including multimodal user behavior modeling [11] and spatiotemporal prediction [12], offer foundational methodologies for scaling and improving recommendation systems.

## III. BACKGROUND

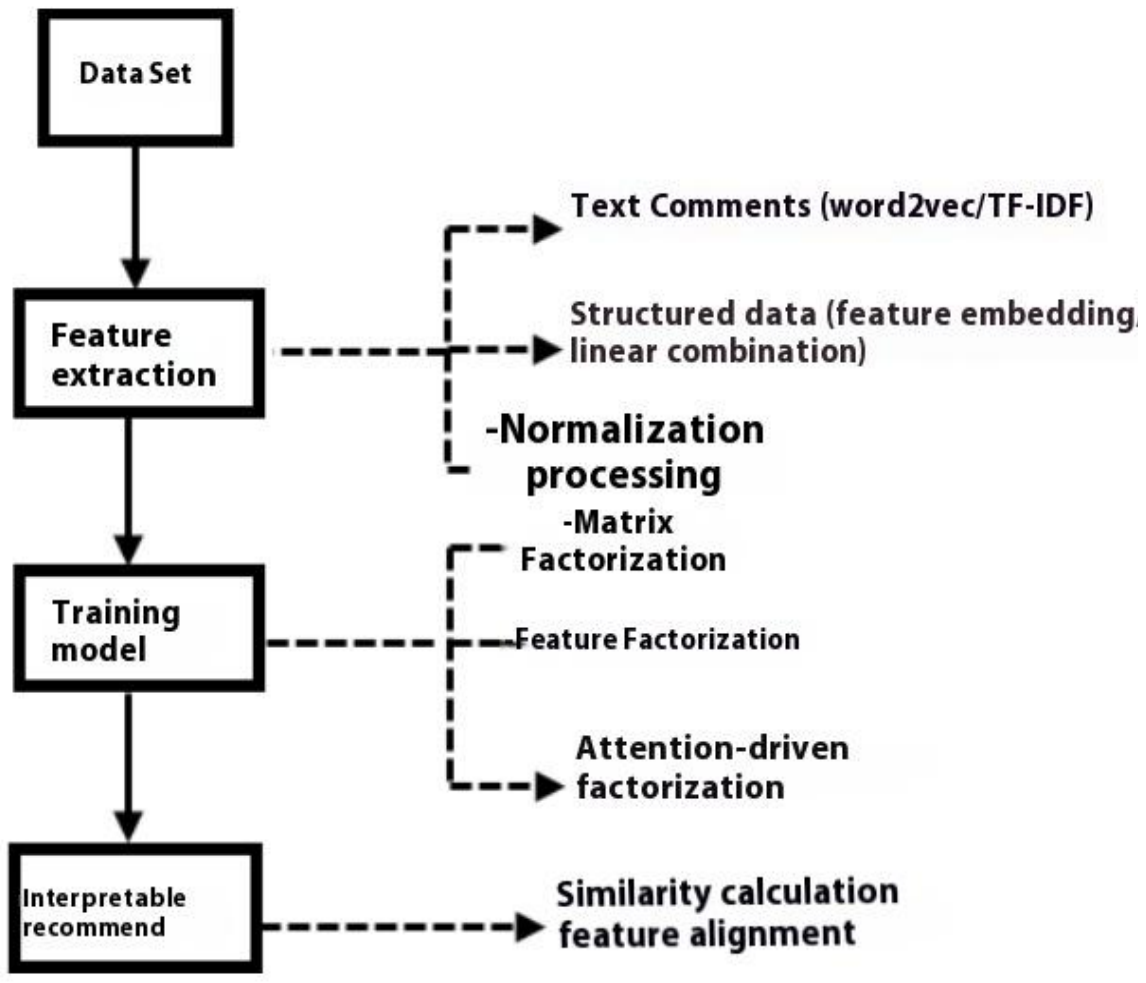

Figure 1 Recommendation Algorithm based on Matrix Factorization

The flow of the matrix factorization-based recommendation algorithm is shown in Figure 1.

Explainable recommendations offer users explanations for recommended results, enhancing user satisfaction and experience. They also provide researchers with accurate grounds for optimizing recommendation models.

### A. *Explainable Recommendation Algorithm Based on Matrix Factorization*

The matrix factorization-based algorithm consists of three main steps, with the specific process as follows:

#### 1) *Construction of Sentiment Dictionary*

Preprocessing operations on original review data involve filtering redundant or interfering information. Item features are extracted, and user sentiment attitudes are determined from users' review text. User emotions are categorized as positive (1) or negative (-1). A sentiment dictionary (item features, sentiment vocabulary, 1 or -1) is constructed to support constructing the item-feature vector matrix.

#### 2) *Matrix Construction*

Users' ratings of items can describe their satisfaction levels. The rating matrix A is constructed based on user ratings.

Attention normalization processing is performed on user's rating for item j, with the specific formula as follows:

$$X_{ij} = \begin{cases} 0, \text{if user } u_i \text{ did not mention feature } F_j \\ 1 + (N-1)\left(\frac{2}{1+e^{-t_{ij}}} - 1\right), \text{else} \end{cases} \quad (1)$$

Given that j represents a feature of item i, if it appears k times in user reviews, it corresponds to k feature/sentiment pairs in the sentiment corpus. Then calculate the average score $s_{ij}$ of k feature/sentiment pairs. Based on this, the corresponding item-feature vector matrix Y can be obtained as follows:

$$Y_{ij} = \begin{cases} 0, \text{if user } p_i \text{ did not mention feature } F_j \\ 1 + \frac{N-1}{1+e^{-k \cdot s_{ij}}}, \text{else} \end{cases} \quad (2)$$

3) *Missing values in the matrix, where users haven't rated items or features, need to be evaluated through loss function optimization.*

The quadratic loss function optimization for matrices *X* and *Y* is shown as follows:

$$\text{minimize}\{\lambda_x \| U_1 V^T - X \| + \lambda_y \| U_2 V^T - Y \|_F^2\} \quad (3)$$

Missing values in rating matrix A are estimated using the same method, considering latent factors affecting ratings. Thus, the minimization loss function for A is:

$$\text{minimize } \{\| PQ^T - A \|_F^2\} \quad (4)$$

After parameter convergence, missing values in *X*, *Y*, and *A* can be obtained. The rating value for user *i* on target *j* is calculated based on the maximum feature index, using the following formula:

$$R_{ij} = \alpha \cdot \frac{\sum_{c \in c_i} \tilde{X}_{lc} \cdot \tilde{Y}_{jc}}{kN} + (1-\alpha) A_{lj} \quad (5)$$

To enable traceability of the recommendation process, item features need to be decomposed, and user sentiments toward features are extracted from the review corpus. Combining feature dimensions and user preferences, matrix factorization can be extended to tensor factorization.

## IV. METHOD

In this research, we analyzed and designed a recommendation algorithm that incorporates Collaborative Filtering (CF) mechanisms based on Graph Neural Networks (GNN), addressing the smoothing problem faced by existing GNN-based recommendation algorithms. The proposed algorithm consists of four main components: an embedding layer, an aggregation propagation layer, a rating prediction layer, and model training.

The embedding layer initializes embeddings for each user and item by incorporating their interaction history. It constructs a normalized adjacency matrix for the graph neural network based on user-item interaction histories, generating initial embedding matrices that capture user-item relationships. In the aggregation propagation layer, collaborative signals between users and items are obtained through initial residual connections and identity mapping. The rating prediction layer aggregates collaborative signals from different layers to generate the model's predicted ratings. Finally, the model is trained through loss function minimization.

### *A. Embedding Layer*

This paper constructs a graph neural network composed of two types of nodes: users and items. A user-item bipartite graph G=(U, I) is constructed, where $U = u_1, u_2, \cdots, u_m$ represents a set of m users, and $I = i_1, i_2, \cdots, i_n$ represents a set of n items. Let $R = (r_{ij})nm$ be the rating matrix representing historical user-item interactions, where elements $r_{ij}$ in R can only have values of 1 or 0. When $r_{ij} = 1$, it indicates that user i has historical interaction with item j; when $r_{ij} = 0$, it indicates no historical interaction between user *i* and item *j*. A user can rate multiple items, and multiple users can rate a single item. Figure 2 illustrates a user-item bipartite graph comprising three users and four items, along with its corresponding rating matrix.

The adjacency matrix A of the user-item bipartite graph G can be expressed as:

$$A = \begin{bmatrix} 0 & R \\ R^T & 0 \end{bmatrix} \quad (6)$$

where $(\cdot)^T$ represents matrix transpose. Subsequently, the degree matrix D is used to normalize matrix A, resulting in the Laplacian matrix $S = D^{-\frac{1}{2}} A D^{-\frac{1}{2}}$.

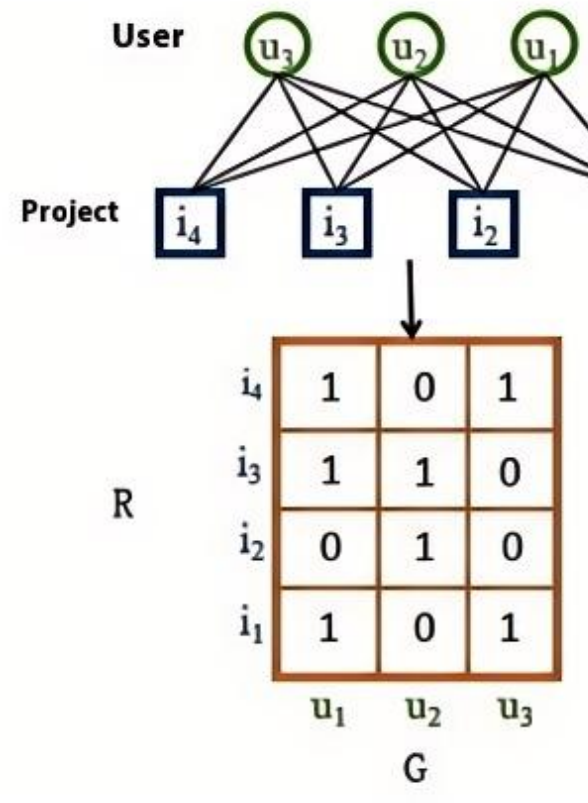

Figure 2 Second Order Phase Diagram and Rating Matrix

The degree matrix D is a diagonal matrix where the diagonal elements d are the sum of all elements in row i of matrix A. Matrix S is called the normalized Laplacian matrix. Generally, matrix S is a large sparse matrix. Direct computation using S would consume significant computational resources. Therefore, we introduce the initial user-item embedding matrix $E^{(0)} = SW$, where $W$ represents the network parameters to be learned. $W$ can be densified to improve computational efficiency. Additionally, $W$ serves as the weight matrix for user and item historical features, enabling better learning of user and item embedding representations.

## B. Aggregation Propagation Layer

The aggregation propagation layer employs a graph convolutional neural network to identify similar nodes in the graph and aggregates information from similar nodes to obtain higher-order collaborative signals for each node. This layer consists of two main components: initial residual connections and identity mapping. The introduction of these components helps reduce the over-smoothing of the graph neural network and aggregates more historical user-item interaction information, leading to more accurate recommendations.

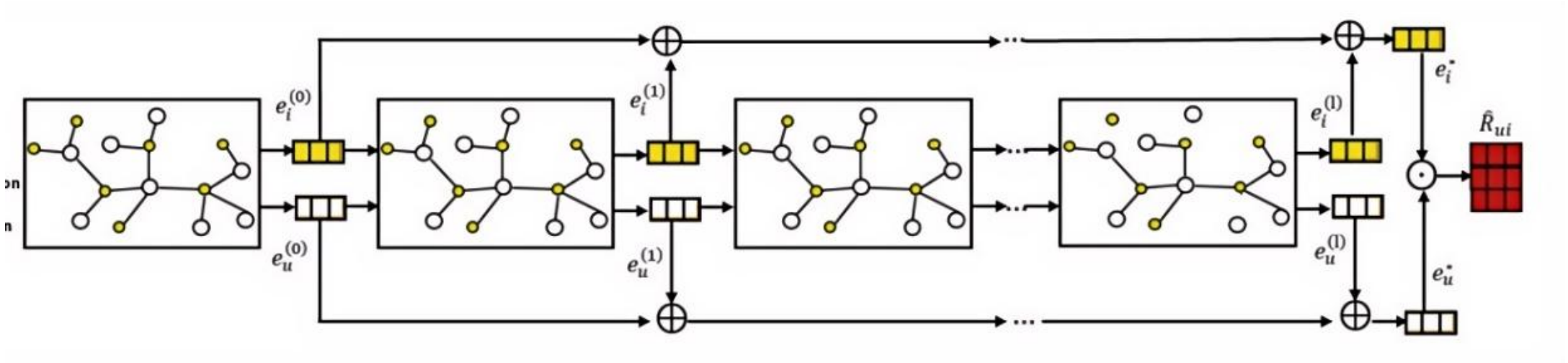

Figure 3: Aggregated Propagation and Rating Prediction Flow Diagram

### 1) Initial Residual Connection

Residual connections learn the differences between layers in the network, enabling the network to continue learning high-order features in deeper layers, thereby improving network performance. The typical form of residual connections in graph neural networks is:

$$E^{(l)} = SE^{(l-1)} + E^{(l-1)} \tag{7}$$

where $E^{(l)}$ represents the output of network layer l, $E^{(l-1)}$ represents the input to layer l, and S is the normalized Laplacian matrix. While residual connections can moderately alleviate over-smoothing in graph neural networks, the phenomenon still occurs when stacking multiple layers. Moreover, using only the results from layer l-1 as input to layer l is not entirely reasonable, as over-smoothing may have already occurred before layer l-1. Therefore, initial residual connections are adopted instead:

$$E^{(l)} = SE^{(l-1)} + E^{(0)} \tag{8}$$

where $E^{(0)}$ represents the initial embeddings of users and items. Initial residuals ensure that the network retains initial information at any layer, preventing information loss and convergence to a subspace in deeper layers [13].

A parameter $\alpha(0<\alpha<1)$ is introduced to adjust the proportion between the previous layer's aggregated propagation features and initial residuals [14], enhancing network flexibility and expressiveness:

$$E^{(l)} = (1-\alpha)SE^{(l-1)} + \alpha E^{(0)} \tag{9}$$

### 2) Identity Mapping

To capture non-linear features between users and items, non-linear activation functions are added to equation. However, using only initial residual connections with activation functions cannot completely prevent over-smoothing [15]. Drawing inspiration from residual networks [16], identity mapping is further incorporated:

$$E^{(l)} = \sigma(((1-\alpha)SE^{(l-1)} + \alpha E^{(0)})(I + W^{(l-1)})) \tag{10}$$

where σ represents the non-linear activation function, I is the identity matrix, and $W^{(l-1)}$ represents trainable parameter matrices.

A parameter $\beta(0 < \beta < 1)$ is introduced to control the proportion between I and $W^{(l-1)}$ to improve model performance. The final expression for the aggregation propagation layer is:

$$E^{(l)} = \sigma(((1-\alpha)SE^{(l-1)} + \alpha E^{(0)})((1-\beta)I + \beta W^{(l-1)})) \tag{11}$$

In equation, the eigenvalues of the identity mapping matrix $(1-\beta)I + \beta W^{(l-1)}$ are controlled by β. An appropriate β value can make the maximum eigenvalue of $(1-\beta)I + \beta W^{(l-1)}$ close to 1, thereby reducing information loss during network propagation [17].

## C. Rating Prediction Layer

Through the embedding layer propagation and l layers of propagation, we obtain user-item matrices $E^{(k)} (0 \le k \le l)$ containing collaborative signals between users and items. Since $E^{(k)}$ from different layers contains different collaborative signals, we need to aggregate the user-item matrices $E^{(k)}$ from different layers:

$$E^{*} = \alpha_0 E^{(0)} + \alpha_1 E^{(1)} + \cdots + \alpha_l E^{(l)} \quad (12)$$

where $\alpha_k (k = 0, 1, \ldots, l)$ are arbitrary real numbers.

Let $\boldsymbol{E}^{(k)} = \begin{bmatrix} e_u^{(k)} \\ e_i^{(k)} \end{bmatrix}$, then $E^{(*)} = \begin{bmatrix} e_u^{*} \\ e_i^{*} \end{bmatrix} = \begin{bmatrix} \sum_{k=0}^{l} a_k e_u^{(k)} \\ \sum_{k=0}^{l} a_k e_i^{(k)} \end{bmatrix}$.

Finally, the model's predicted ratings can be obtained through $E^{*}$ as follows:

$$\hat{R}_{ui} = (e_u^{*})^{T} e_i^{*} \quad (13)$$

This layered approach allows the model to combine collaborative signals from different propagation depths [18], leading to more comprehensive and accurate rating predictions. The final prediction is computed as the inner product of the aggregated user and item embeddings, capturing the overall similarity between users and items across all layers of the network.

Let $\boldsymbol{E}^{(k)} = \begin{bmatrix} e_u^{(k)} \\ e_i^{(k)} \end{bmatrix}$, then $E^{(*)} = \begin{bmatrix} e_u^{*} \\ e_i^{*} \end{bmatrix} = \begin{bmatrix} \sum_{k=0}^{l} a_k e_u^{(k)} \\ \sum_{k=0}^{l} a_k e_i^{(k)} \end{bmatrix}$.

Finally, the model's predicted ratings can be obtained through $E^{*}$ as follows:

$$\hat{R}_{ui} = (e_u^{*})^{T} e_i^{*} \quad (14)$$

This layered approach allows the model to combine collaborative signals from different propagation depths, leading to more comprehensive and accurate rating predictions. The final prediction is computed as the inner product of the aggregated user and item embeddings, capturing the overall similarity between users and items across all layers of the network.

# V. Experiment

## A. Dataset

This chapter utilizes three datasets, namely the Gowalla dataset, the Yelp-2018 dataset, and the Amazon-Book dataset. The Gowalla dataset is a check-in dataset, where the locations shared by users from the Gowalla dataset are considered as items in this chapter. The Yelp-2018 dataset originates from the 2018 challenge, where local business establishments from this dataset are regarded as items. The Amazon-Book dataset is derived from the Amazon shopping website, and this chapter selects books sold on the Amazon shopping website as items. This chapter selects 10-core operations to ensure the quality of the datasets in the experiments, which means that each selected node has at least 10 interactions. In other words, each user has at least 10 item interaction records, and each item also has at least 10 user interaction records.

## B. Baselines

This chapter compares five mainstream algorithms with the algorithm presented in this chapter.

BPRMF [19], GCMC [20], NGCF[21], LR-GCCF[22], LGC[23]. Among the five comparative algorithms mentioned above, the first algorithm, BPRMF, is a classic matrix factorization recommendation algorithm [24], while the latter four are recommendation algorithms based on graph neural networks.

## C. Comparative Experiment

To thoroughly validate the recommendation performance of the graph neural network-based recommendation algorithm proposed in this chapter, experiments were conducted comparing it with five mainstream recommendation algorithms, with the results shown in Table 1. From these results, we can draw the following conclusions:

Table 1: The Results of Performance Comparison

| Dataset | Metrics | BPRMF | GCMC | NGCF | LGC | Ours |
|---|---|---|---|---|---|---|
| Gowalla | Recall@20 | 0.1265 | 0.1447 | 0.1499 | 0.1754 | 0.1771 |
| | NDCG@20 | 0.1087 | 0.1181 | 0.1287 | 0.1488 | 0.1491 |
| Yelp-2018 | Recall@20 | 0.0424 | 0.0453 | 0.0567 | 0.0626 | 0.0632 |
| | NDCG@20 | 0.0347 | 0.0371 | 0.0467 | 0.0515 | 0.0521 |
| Amazon-Book | Recall@20 | 0.0245 | 0.0347 | 0.0337 | 0.0388 | 0.0409 |
| | NDCG@20 | 0.0192 | 0.0265 | 0.0248 | 0.0292 | 0.0325 |

1): The matrix factorization-based algorithm BPRMF's recommendation performance is low because it predicts ratings between users and items without considering the graph structure between users and items, or the interaction information.

2): The five-graph neural network-based algorithms (GCMC, NGCF, LR-GCCF, LGC, and this chapter's algorithm) outperform BPRMF. These algorithms use graph neural networks to exploit graph structure information, including explicit user-item interaction data, enabling a more comprehensive representation of user-item features based on historical connectivity.

3): The graph neural network-based recommendation algorithm proposed in this chapter outperforms the other four graph neural network-based algorithms (GCMC, NGCF, LR-GCCF, and LGC) in terms of recall rate and normalized discounted cumulative gain across the three datasets. This is due to the use of initial residual connections in the aggregation propagation layer, which retains initial information and avoids model convergence to a subspace. Additionally, the algorithm adds identity mapping in the aggregation propagation layer, enabling the network to learn nonlinear relationships between users and items.

## VI. Conclusion

This paper introduces an innovative deep graph neural network-based recommendation algorithm specifically designed to address the persistent challenges of over-smoothing and interpretability within GNN-based recommendation systems. By incorporating initial residual connections and identity mappings into the aggregation and propagation stages, the proposed model preserves critical user-item interaction details across layers, enabling the network to maintain meaningful signal propagation even as its depth increases. This approach mitigates the adverse effects of over-smoothing—where information becomes indistinguishable in deeper layers—thereby enhancing recommendation accuracy and allowing for a more nuanced representation of user preferences. Additionally, the model emphasizes interpretability, an increasingly vital component in recommendation systems, ensuring that the suggestions provided are both transparent and aligned with dynamic user behavior. By synergizing collaborative filtering techniques with GNN architectures, this work not only advances the technical robustness of recommendation algorithms but also addresses key user-centric demands, creating a system that adapts flexibly to shifts in user interests. Overall, this research contributes a comprehensive framework for building explainable, high-performing recommendation systems capable of scaling within the complex and data-intensive environments of modern online platforms, thus offering valuable insights and tools for future development in the field of personalized recommendations.